\documentclass[final,3p]{elsarticle}

\usepackage{xcolor}
\usepackage{hyperref}
\hypersetup{
    colorlinks=true,
}

\usepackage[T1]{fontenc} 
\usepackage{graphicx}
\usepackage{amsmath,amssymb}
\usepackage{booktabs}
\usepackage{subcaption}
\usepackage[nice]{nicefrac}

\journal{Computer Aided Geometric Design (GMP 2026)}

\graphicspath{{/}{figures/}}

\newcommand{\rem}[1]{}

\newcommand{\matlab}{\textsc{Matlab}}

\begin{document}
\begin{frontmatter}

\author[a1,a2]{Stefan Pillwein}
\author[a2]{Alexander Hentschel}
\author[a2]{Markus Lukacevic}
\author[a1]{Przemyslaw Musialski}

\affiliation[a1]{organization={New Jersey Institute of Technology},
city={Newark, NJ},
country={United States}}
\affiliation[a2]{organization={Technische Universit\"{a}t Wien},
city={Vienna},
country={Austria}}

\title{Geometric Guidance for Globally Synchronized Deployment of Elastic Geodesic Grids}

\begin{abstract}
Elastic geodesic grids deploy from flat to spatial configurations via complex nonlinear motion that is difficult to represent robustly for simulation. We present a geometric guidance framework that discretizes deployment as synchronized, time-coupled deformation trajectories. Starting from inverse tracing---collapsing the deployed structure with a lightweight rod model while recording node paths under a shared parameter---we obtain feasible node paths and formulate a polyline approximation problem that selects {globally synchronized} time steps and minimizes a robust tail-aggregated deviation measure under monotonicity constraints. {We solve the resulting non-smooth optimization problem via global optimization to obtain compact, synchronized displacement sequences for all paths simultaneously}. We evaluate the method using geometry-centric metrics (deviation versus step count, scaling with trajectory count) and demonstrate its utility by driving finite element deployment simulations that avoid intermediate buckling and capture deployment-induced prestress. 
\end{abstract}


\end{frontmatter}

\section{Introduction}
\label{sec:introduction}

\begin{figure*}[!h]
  \centering
  \begin{minipage}[t]{0.49\textwidth}
    \centering
    \includegraphics[width=0.88\columnwidth]{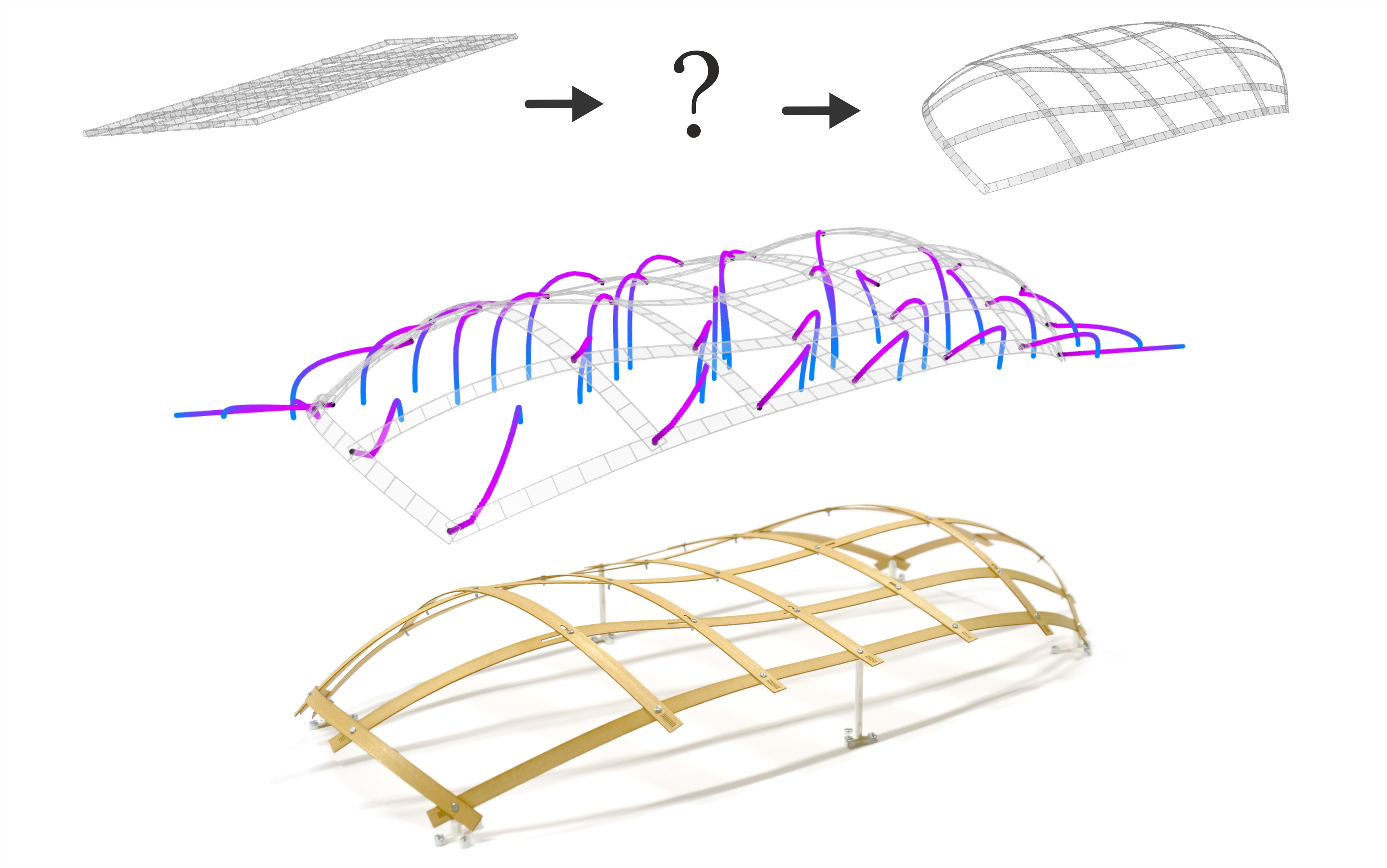}
    \caption{Deploying elastic gridshells involves non-linear motion that can enter buckled configurations. We compute globally synchronized guidance paths and discretize them into piecewise-linear displacement sequences for FEA simulations.} 
    \label{fig:overview}
  \end{minipage}\hfill%
  \begin{minipage}[t]{0.49\textwidth}
    \centering
    \includegraphics[width=0.95\columnwidth]{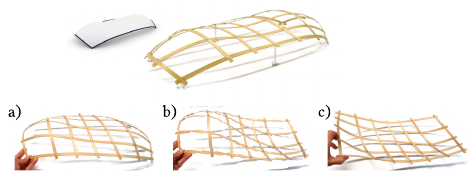}
    \caption{
      Elastic gridshells may admit multiple stable equilibria under deployment.
      (a) Intended deployed configuration.
      (b,c) Alternative buckled equilibria resulting from geometric ambiguity during actuation.
    }
    \label{fig:buckling_grids}
  \end{minipage}
\end{figure*}

Deployable elastic structures transition from a planar state to a spatial target state through stored elastic energy. These systems allow for lightweight constructions that realize complex curved configurations from flat manufacturing layouts, utilized in applications such as adaptive facades, prestressed bridges, and deployable shelters (Fig~\ref{fig:overview}). Recent work in geometric modeling has developed computational methods for designing such structures, including scissor-like deployable grids and elastic geodesic grids \cite{Pillwein2020, Pillwein2020a, pillwein2021, Pillwein2021b,Soriano2019, Haskell21, Panetta2019}.

While form-finding methods provide initial geometric candidates, structural assessment requires detailed finite element analysis (FEA).  {Accurate deployment simulation is specifically needed to capture the internal stresses that accumulate during erection: these deployment-induced forces produce a prestressed initial state that governs the subsequent load-bearing capacity and stiffness of the deployed structure through stress-stiffening effects.} A central discrepancy exists between the geometric definition of deployment and the input requirements of standard solvers: the physical deployment follows continuous, curved trajectories in space, but FEA workflows typically prescribe boundary conditions as sequences of linear displacement steps. Mismatches between the discretized linear input and the natural curved trajectory introduce undesired axial compression in the slender grid elements. This undesired compression can trigger local buckling and numerical non-convergence, preventing the simulation of the deployed state.  {Most importantly, the
discretization cannot be performed independently for each anchor trajectory:
all paths share a single time parameterization imposed by the mechanism's
kinematic coupling, making the problem a joint optimization over all paths
simultaneously rather than independent per-curve resampling.}

This paper addresses deployment guidance as a geometric representation and discretization problem. Our goal is to guide selected anchor points of the structure--- {which include both boundary nodes and internal lattice intersection points used as kinematic control locations in the finite element model}---along synchronized 3D trajectories parametrized by a shared scalar value, and to convert these trajectories into solver-compatible, multi-step displacement boundary conditions. We acquire feasible guidance trajectories by inverse tracing on a simplified discrete elastic rod model. We then discretize these paths by solving a joint polyline approximation problem under monotonicity constraints. The resulting displacement sequences minimize deviation from the natural trajectory while maintaining synchronization, serving as robust kinematic inputs for quasi-static simulations.

Our contributions are the following:
\begin{itemize}
\item We describe a workflow to acquire coupled 3D deployment trajectories via inverse tracing on a reduced-order proxy model.
\item We formulate synchronized path discretization as a constrained polyline approximation problem  {across multiple trajectories} that controls a robust assembly-deviation metric while preserving a shared parameterization.  {The resulting non-smooth optimization problem is solved via global optimization over the shared parameter vector.} 
\item We validate the geometric representation by applying the computed displacement sequences as boundary conditions in commercial finite element software, demonstrating successful deployment of tested examples with non-trivial topologies.
\end{itemize}

The organization of the paper is as follows: Section~\ref{sec:related} reviews geometric discretization and deployable structures; Section~\ref{sec:challenges} details the geometric mismatch in deployment simulation; Section~\ref{sec:geometry_wrapping} presents the trajectory acquisition and synchronized discretization method; and Section~\ref{sec:discussion} discusses limitations and Section~\ref{sec:conclusions}  concludes the work. Implementation details for the validation setup are provided in~\ref{sec:abaqus}.

\section{Related Work}
\label{sec:related}

\paragraph{Elastic Gridshells}
Elastic gridshells are bending-active structures that realize spatial configurations from initially planar layouts through elastic deformation of slender elements. Early architectural realizations include Shukhov’s Rotunda of the Panrussian Exposition \cite{Shukhov1896} and the Multihalle at the Mannheim Bundesgartenschau by Frei Otto \cite{Happold1975}. More recent built examples include the Downland Gridshell \cite{downland,Harris2003} and the Ephemeral Cathedral \cite{peloux2016}. These structures combine shell-like load transfer with material-efficient grid layouts \cite{adriaenssens2014shell}, and their geometry emerges from bending, twisting, and boundary constraints applied during deployment. Broader context for bending-active structures and built gridshell practice is provided by~\cite{Lienhard2013} and~\cite{Schling2018}. 

Deployment induces internal stresses that influence the realized equilibrium shape and subsequent load response. Long-term material effects such as creep and relaxation can further alter the geometry over time, as observed in the Multihalle \cite{peloux2016}. These effects are time-consuming to investigate experimentally \cite{lara2018}, but have been studied in simulation through models linking wood deformation to hygroscopic processes \cite{gronquist2020,Autengruber2020,Autengruber2021}. In this work, we focus on the geometric kinematics of the initial deployment phase rather than long-term material evolution.

\paragraph{Elastically Deployable Structures}
Using elasticity to generate curved shapes from flat states appears in many geometric and fabrication contexts. Free-form surfaces can be approximated by assemblies of developable surfaces \cite{Stein18,Binninger21}, which are subsequently bent into shape. Curved folding provides another mechanism to realize curved configurations from planar sheets \cite{Tang16,Kilian2008,Kilian2017,Rabinovich19}. Closely related are approaches that explore approximately isometric shape changes \cite{Jiang20}, emphasizing bending deformations without stretching.

Beyond developability, auxetic structures exploit negative Poisson ratios to enable expansion \cite{Konakovic2016,Konakovic-Lukovic2018,Panetta2021}, and bi-stable elastic configurations have been investigated in this context \cite{Chen21}. Deforming and combining elastic meso-materials yields additional pathways for generating curved forms from flat cellular designs \cite{Malomo2018a,laccone2019flexmaps,Laccone2021}. Programmable elastic structures actuate planar layouts into free-form shapes \cite{Guseinov2017} or evolve into doubly curved surfaces over time \cite{Guseinov2020}. Weaving approaches based on thin strips are also related, ranging from geodesic weaves \cite{Vekhter2019} to weaves with curved strips \cite{Ren21}.

\begin{figure*}[!t]
  \centering
  \begin{minipage}[t]{0.49\textwidth}
    \centering
    \includegraphics[width=0.95\columnwidth]{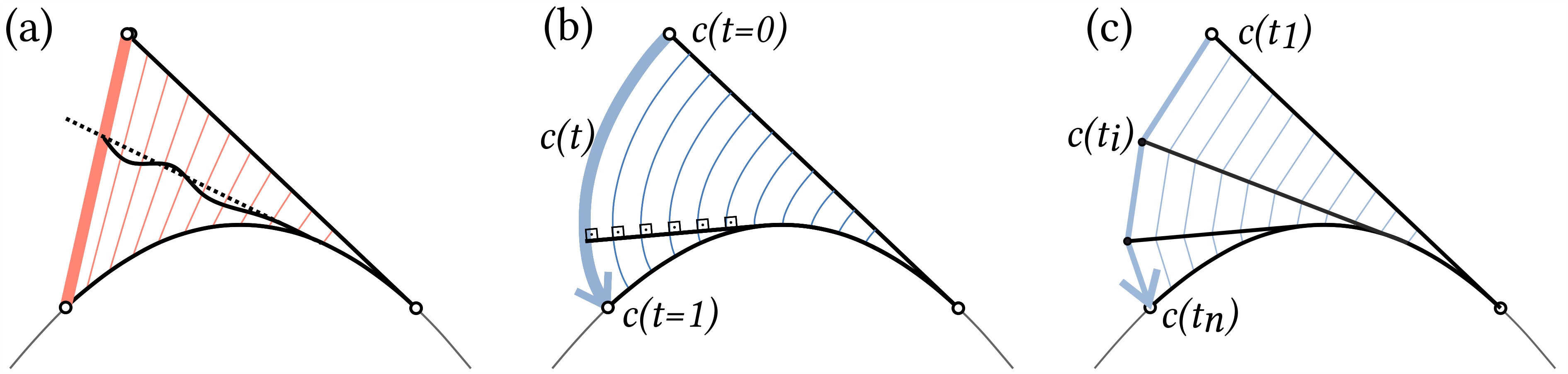}
    \caption{Geometry of wrapping a straight elastic element along a curve. Wrapping elements along linear paths using a single step leads to compression and severe buckling (left). The involute $c(t)$ (cf. Section \ref{sec:single_path}) yields stretch-free curved paths (center), where $t$ is the time parameter of the deformation. To adopt them for computational usage, they need to be linearized (right), keeping the compression of the elastic element low.}
    \label{fig:involute}
  \end{minipage}\hfill%
  \begin{minipage}[t]{0.49\textwidth}
    \centering
    \includegraphics[width=\columnwidth]{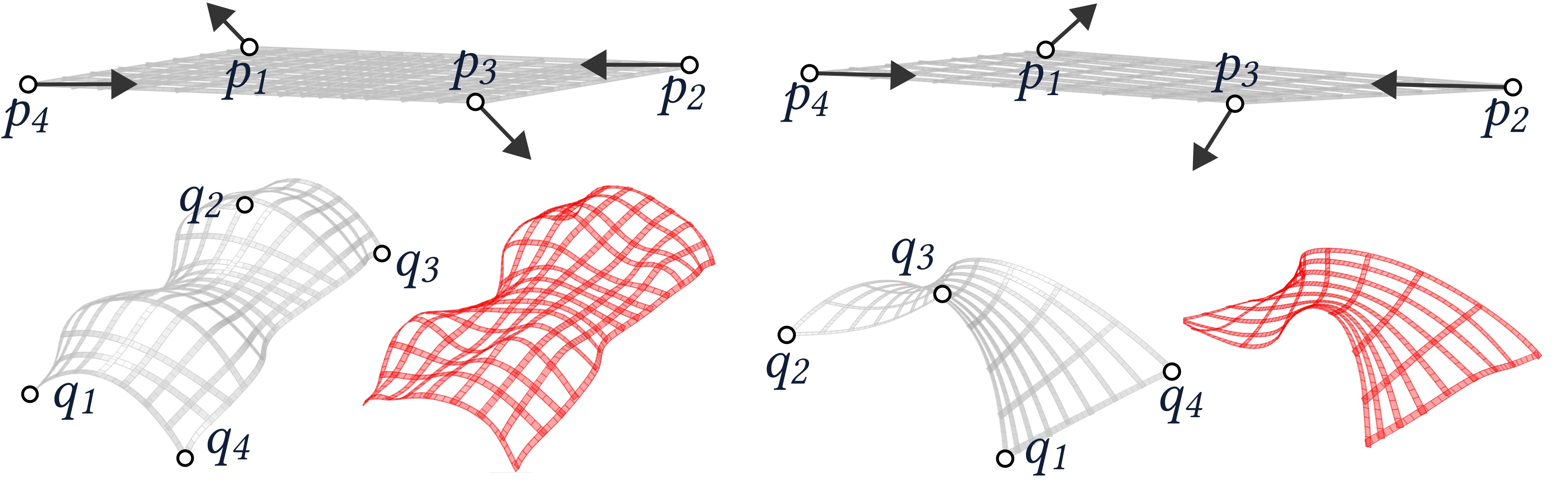}
    \caption{ Linearly moving the corners of an elastic geodesic grid to the final locations in space fails: The grid may move into an unintended local minimum of elastic energy (i.e. adopt the wrong equilibrium shape). The results were computed using simplified simulation techniques (cf. Section \ref{sec:related}), which deal well with buckling. However, it poses serious problems for FEA models with up to hundreds of thousands of variables.}
    \label{fig:deployment_fail}
  \end{minipage}
\end{figure*}

\paragraph{Scissor-like Deployable Gridshells}
A class of deployable elastic structures closely related to our work employs scissor-like mechanisms. These systems have been studied in computer graphics \cite{Panetta2019,Pillwein2020,Vekhter2019,Ren21}, engineering \cite{Haskell21}, and design \cite{Soriano2019}. They can be fabricated and assembled in a compact, planar state and deployed into spatial configurations through built-in kinematic mechanisms, reducing the need for scaffolding during erection.

X-Shells \cite{Panetta2019,X_shells_pavilion} employ straight or curved elements connected by rotational one degree-of-freedom joints and are actuated through joint rotation. Other approaches use straight lamellae with rotational joints to approximate target surfaces \cite{Soriano2019,Haskell21}. Elastic geodesic grids (EGG) \cite{Pillwein2020,Pillwein2020a,pillwein2021,Pillwein2021b} extend this idea by employing sliding connections with one rotational and two translational degrees of freedom (see Figure~\ref{fig:notches_abstract}). These grids behave as one-degree-of-freedom mechanisms and deploy by changing an internal angle or by pulling at selected boundary points.

\paragraph{Simulation Techniques for Form-Finding and Deployment}
Form-finding simulations for elastic structures often rely on geometric simplifications and reduced-order physical models to obtain robust feedback during design. Many approaches compute equilibrium configurations directly from geometric initializations without explicitly simulating the deployment path. Typical discretizations represent elements as polylines with assigned cross-sections. Common techniques include geometric energy formulations such as the Discrete Elastic Rods model \cite{Bergou2008,Bergou2010} and force-based approaches that minimize out-of-balance forces \cite{dAmico2015,Lefevre2017,Sakai2020}. Incorporating joint constraints usually requires additional constraint terms, resulting in hybrid geometric–physical formulations.

While such methods are effective for design exploration, they do not inherently provide synchronized, solver-compatible actuation trajectories for detailed deployment simulation. 
The geometric origin of this mismatch, illustrated for a single elastic element by involute trajectories, is shown in Figure~\ref{fig:involute}. 
The geometric discretization of deployment paths addressed in this work complements these approaches by explicitly representing the kinematic evolution required for validation-stage simulations.

\section{Geometric Failure Modes of Deployment}
\label{sec:challenges}

\begin{figure*}[!t]
  \centering
  \begin{minipage}[t]{0.49\textwidth}
    \centering
    \includegraphics[width=\columnwidth]{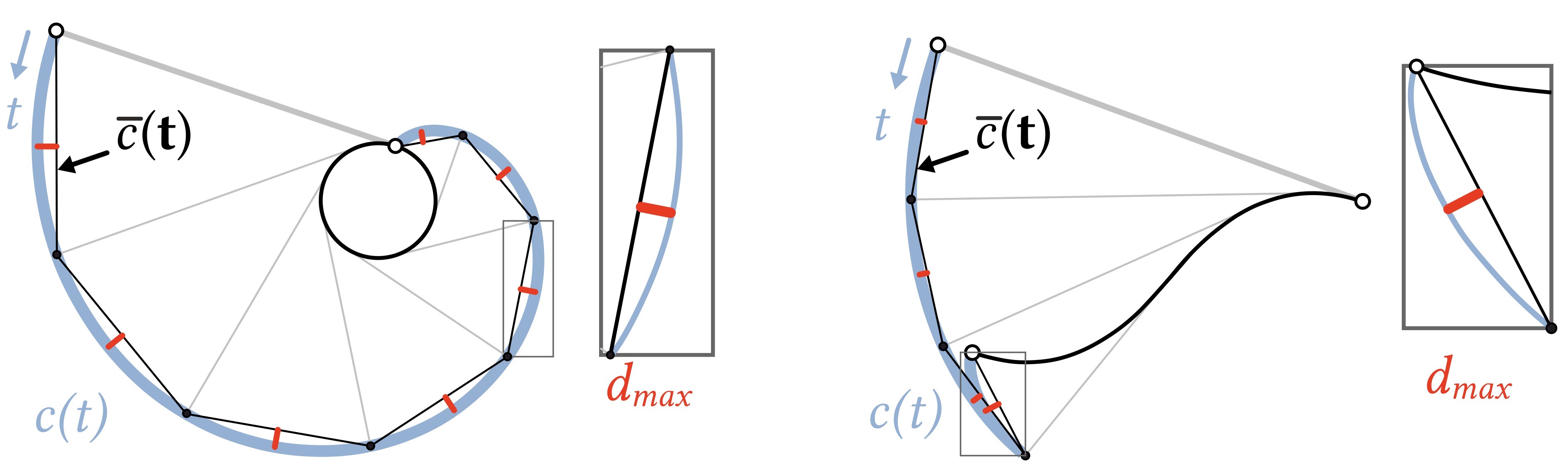} %
    \caption{Examples of straight elements (grey) are wrapped on target curves (black). 
    The right example shows a cusp in its involute (blue).
    {For every line segment of $\overline{c}(\mathbf{t})$, we identify the per-segment deviations $d_i$ and finally the overall biggest deviation $d_{max} = \text{max}(\,[d_1,...d_n]\,)$ for the whole polyline $\overline{c}(\mathbf{t})$.}}
    \label{fig:involute_examples}
  \end{minipage}\hfill%
  \begin{minipage}[t]{0.49\textwidth}
    \centering
    \includegraphics[width=\columnwidth]{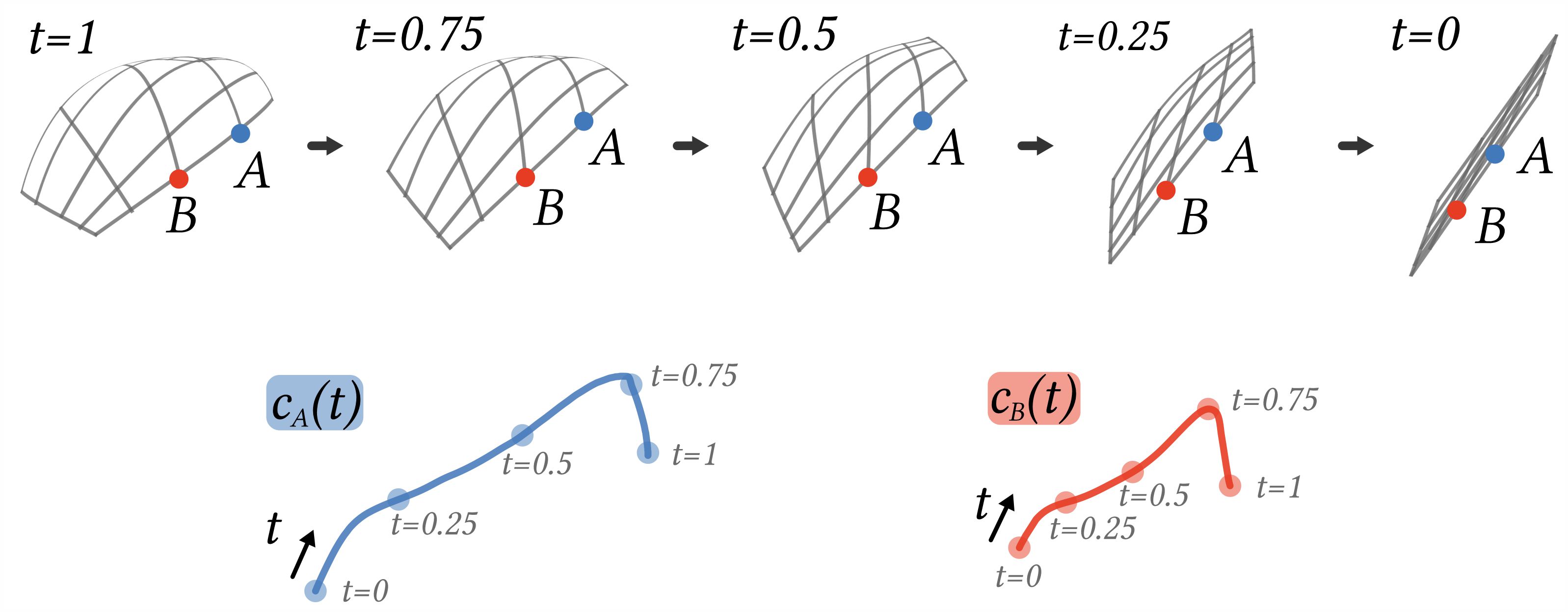}
    \caption{ The computation of deployment paths for a scissor-like gridshell. We collapse the deployed grid in a high number of steps and trace the movement of points. The resulting polylines for points $A,B$ are depicted below. The geometrically complicated movement of the structure along these polylines is parametrized by a globally synchronized time parameter $t$.}
    \label{fig:deployment_paths}
  \end{minipage}
\end{figure*}

The fundamental difficulty in simulating the deployment of elastic gridshells lies in a discrepancy between the continuous nature of the physical mechanism and the discrete, linear input requirements of numerical solvers. We analyze this failure mode as a geometric path discretization problem. Representative inverse-traced deployment trajectories are illustrated in Figure~\ref{fig:deployment_paths}.

\paragraph{Kinematic Incompatibility}
Ideally, the deployment of a mechanism follows a continuous trajectory $\mathbf{x}(t)$ in configuration space that maintains metric consistency (i.e., preserving element lengths and minimizing strain energy). For a specific anchor point $i$, the ideal path is a spatial curve $\mathbf{c}_i(t)$.
However, standard finite element solvers operate in discrete load steps. To move an anchor from state $t_k$ to $t_{k+1}$, the solver imposes a displacement boundary condition $\Delta \mathbf{u}_i = \mathbf{c}_i(t_{k+1}) - \mathbf{c}_i(t_k)$. This prescribes a linear translation along the chord connecting the two points.
Physically, this linear path is generally shorter than the trajectory segment of the deployment trajectory ($||\Delta \mathbf{u}_i|| < \text{length}(\mathbf{c}_i|_{t_k}^{t_{k+1}})$).

\paragraph{Induced Compression and Instability} 
This geometric shortening has severe physical consequences. By forcing the anchor point to travel along the chord, the boundary condition effectively imposes an axial compression on the connected structural members proportional to the path deviation.
Because deployable gridshells are composed of highly slender elements (high aspect ratio), their critical buckling load $P_{crit}$ is very low. Even small geometric deviations $\epsilon$ induced by linear discretization can generate axial forces $F > P_{crit}$.
This drives the system into a bifurcation point where the structure creates a ``local buckle'' to resolve the excess length. Once buckled, the stiffness matrix becomes singular or indefinite, typically causing the numerical solver to diverge or the structure to snap into an invalid energy minimum (as shown in Figures~\ref{fig:buckling_grids} and~ {\ref{fig:deployment_fail}}) from which it cannot recover.  {Figure~\ref{fig:deployment_fail} illustrates this failure concretely: linearly moving the grid corners to their final positions in a single step causes the structure to adopt an unintended buckled equilibrium---a failure mode that resists recovery and prevents any subsequent structural analysis.}

\paragraph{Synchronization Failure}
Furthermore, the deployment is a coupled motion; the trajectory of anchor $i$ is kinematically linked to anchor $j$. Independent discretization of these paths that does not account for a shared deployment parameter destroys this coupling. If one boundary condition advances "faster" than another relative to the mechanism's natural flow, it introduces shear and warping forces that lock the mechanism.
Therefore, a valid deployment simulation requires a \textit{synchronized discretization}: a set of piecewise-linear trajectories that bound the geometric deviation below the critical buckling threshold while maintaining global kinematic coordination.

\section{Geometric Guidance}
\label{sec:geometry_wrapping}

\begin{figure*}
    \centering
	    \includegraphics[width=\textwidth]{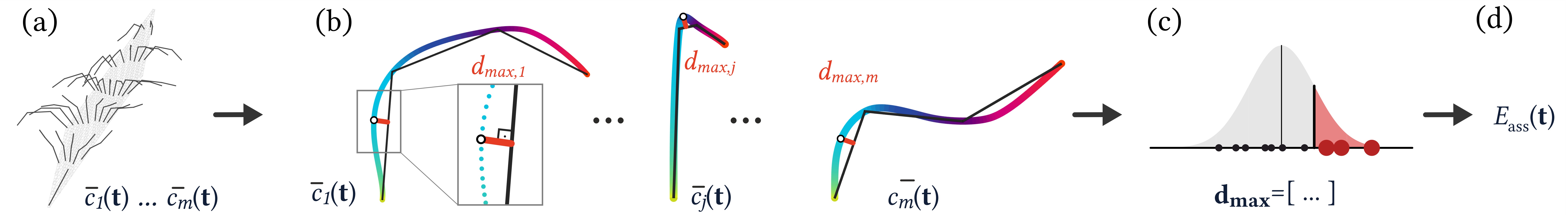} %
	    \caption{
     Our pipeline to linearize the deployment paths:
     (a) We compute low-resolution polylines 
     $\overline {c}_1(\mathbf{t})\,...\, \overline {c}_m(\mathbf{t})$ (black) for all deformation paths $c_1(t)\,...\, c_m(t)$ (colored).
     (b) For each low-resolution polyline  $\overline {c}_j(\mathbf{t})$, we find the biggest deviation $d_\text{max,j}$ to the corresponding deformation path (where the color refers to the parameter $t$).
     (c) From the set of $\mathbf{d}_{\text{max}} = [d_{\text{max},1} \,...\,d_{\text{max},m}]$, we identify the worst and (d) formulate a geometric energy to minimize these deviations.     
     }
	\label{fig:linearize}
\end{figure*}

Our method addresses the deployment simulation problem by constructing a geometric representation of the deployment trajectories that is compatible with finite element solvers. The pipeline consists of two stages: (1) acquiring continuous deployment paths via an inverse tracing operator, and (2) discretizing these paths into synchronized, piecewise-linear displacement sequences that minimize geometric deviation. Figure~\ref{fig:linearize} summarizes the synchronized discretization pipeline.

\subsection{Trajectory Acquisition via Inverse Tracing}
\label{sec:inverse_tracing}

The first step is to obtain the ideal continuous trajectories of the structure's anchor points. Because forward deployment simulation is unstable (as described in Section~\ref{sec:challenges}), we employ an \textit{inverse tracing} strategy. We assume the existence of a known target deployed state (the "design shape") and a flat fabrication state.

We model the structure using a lightweight Discrete Elastic Rods (DER) framework \cite{Bergou2008, Bergou2010}. This reduced-order model is sufficient to capture the kinematic path of minimal energy. We initialize the simulation in the deployed state and add geometric penalty energies to constrain selected grid points to spatial anchors. We then quasi-statically relax these constraints by linearly reducing their weight, guiding the structure toward its planar rest configuration. This "collapse" simulation is numerically stable because it moves from a high-energy state to a low-energy state.

During this process, we record the spatial positions of all $K$ anchor points. This yields a set of dense, time-parameterized curves $\mathcal{C} = \{\mathbf{c}_1(t), \dots, \mathbf{c}_K(t)\}$, where $t \in [0, 1]$ is a shared evolution parameter representing the progression from flat ($t=0$) to deployed ($t=1$). These curves serve as the continuous geometric input for the discretization stage.

\subsection{Single-Path Discretization}
\label{sec:single_path}

Standard FEA solvers require boundary conditions defined as linear displacement steps. For a single curved trajectory $\mathbf{c}(t)$, we must find a sequence of parameter values $0 = t_0 < t_1 < \dots < t_N = 1$ that defines a polyline approximation.

 {The geometric basis for the discretization error is most clearly seen in the involute of a curve. An involute is the locus traced by the endpoint of a taut, inextensible string of fixed length as it unwinds from a base curve; by construction, the string always remains tangent to the base curve, so the resulting path is arc-length-preserving. For an elastic element being wrapped along a target arc, its endpoint therefore traces an involute that keeps the element free of axial strain. This two-dimensional analysis motivates the discretization problem: linearizing such a path---replacing the smooth involute by a chord---introduces a shortfall between the chord length and the arc length, and this shortfall manifests as axial compression in the physical member (Figure~\ref{fig:involute}). The actual deployment paths of a three-dimensional elastic geodesic grid are not analytically constrained to be involutes; they are obtained by inverse tracing on the full reduced-order model (Section~\ref{sec:inverse_tracing}). The involute therefore serves as a geometric illustration of \emph{why} chord-arc deviation causes compression, not as a parametric model of the 3D assembly trajectories.}
We define the deviation cost $E_{dev}$ for a single step as the maximum squared Euclidean distance from the chord to the curve segment:
\begin{equation}
    E_{dev}(t_i, t_{i+1}) = \max_{\tau \in [t_i, t_{i+1}]} \text{dist}(\mathbf{c}(\tau), \text{seg}(\mathbf{c}(t_i), \mathbf{c}(t_{i+1})))^2, 
\end{equation}
where $seg(a,b)$ denotes the straight line segment between points $a$ and $b$. 
Minimizing this deviation is critical because the difference between arc length and chord length manifests as axial compression in the physical structure. For a fixed number of steps $N$, the discretization problem is to find the sequence $\{t_i\}$ that minimizes the maximum deviation over the entire path.

\subsection{Synchronized Discretization}
\label{sec:synchronized_discretization}

Discretizing each anchor trajectory independently would yield asynchronous
parameter values, breaking the kinematic coupling of the mechanism. We
therefore formulate a \textit{synchronized discretization} problem: find a
single global parameter sequence $\mathcal{T} = \{t_0, \dots, t_N\}$,
applied simultaneously to all $K$ trajectories, that minimizes the geometric
mismatch between the piecewise-linear approximation and the continuous paths.

 {Per-curve methods such as curvature-based adaptive sampling cannot
serve as an alternative. Such methods select a locally optimal sequence
$\mathcal{T}_j$ for each path~$\mathbf{c}_j(t)$, tuned to that path's
individual curvature profile. However, elastic geodesic grids behave as
one-degree-of-freedom mechanisms (Section~\ref{sec:related}): every anchor
point is kinematically coupled to every other through the grid's internal
constraints. Prescribing asynchronous boundary condition sequences---even
individually well-approximated ones---forces some anchors to lead or lag the
mechanism's natural configuration flow, introducing artificial shear and
warping that locks the grid or drives it into an unintended equilibrium. The
shared global sequence~$\mathcal{T}$ is therefore not a heuristic
simplification but a structural requirement imposed by the mechanism's
coupling.}

\paragraph{Error Metric}
For a given $\mathcal{T}$, let $d_{\max,j}$ be the maximum point-to-chord
deviation on trajectory $j \in \{1,\dots,K\}$. We aggregate over the top
16\% of deviations (indices $\mathcal{S}$ with values exceeding
mean$+1\sigma$):
\begin{equation}
E_{\text{ass}}(\mathcal{T}) \;=\;
\frac{1}{|\mathcal{S}|} \sum_{j \in \mathcal{S}} (d_{\max,j})^2.
\end{equation}
 {This threshold is a standard robust compromise: a pure maximum is
sensitive to single outlier trajectories, while a plain mean can mask the
worst-case anchors that govern buckling risk; the mean$+1\sigma$ tail
concentrates the objective on the structurally critical subset without being
driven by noise.}

\paragraph{Optimization Problem}
We minimize $E_{\text{ass}}$ over all shared monotone sequences:
\begin{equation}
\min_{t_1,\dots,t_{N-1}}\; E_{\text{ass}}(\mathcal{T})
\quad\text{s.t.}\quad
0 < t_1 < \cdots < t_{N-1} < 1,
\label{eq:opt_assemblies}
\end{equation}
with $t_0=0$ and $t_N=1$ fixed.  {The uniform subdivision
$t_i = i/N$ is a feasible point of this problem---one admissible shared
monotone sequence. Consequently, any global minimizer of
Eq.~\eqref{eq:opt_assemblies} cannot perform worse than uniform spacing for
the same $N$ on $E_{\text{ass}}$. \ref{app:uniform-suboptimal}
provides an auxiliary asymptotic argument for a related minimax chord--arc
criterion that gives geometric intuition for why non-uniform spacing is
preferred.}

\paragraph{Solver}
 {The objective $E_{\text{ass}}$ has a nested maximum structure: for
each candidate $\mathcal{T}$, it evaluates the maximum point-to-chord
deviation along each of the $m$ paths and then aggregates the high-deviation tail
across the assembly. This max-of-max structure admits no useful gradient, and
the coupling of all paths through a shared $\mathcal{T}$ creates numerous
local minima in which gradient-based solvers reliably stall.} We therefore
employ a Genetic Algorithm (GA)~\cite{Goldberg1989} {, implemented in
\matlab{} using the built-in \texttt{ga} solver}.  {GA is run
with $\texttt{PopulationSize}=50$, $\texttt{MaxGenerations}=20$,
$\texttt{MaxStallGenerations}=5$, $\texttt{FunctionTolerance}=10^{-10}$; all
remaining options use MATLAB defaults.}

\begin{figure*}[!b]
  \centering
  \begin{minipage}[t]{0.49\textwidth}
  \vspace{0pt}
    \centering
    \footnotesize
    \captionof{table}{
      Quantitative results of our method. We measure the approximation error in terms of our objective $E_{\text{ass}}$, the number of deformation paths $m$, and the required number of linear segments $N$. The timings $t_{opt}$ to solve the Optimization Problem \eqref{eq:opt_assemblies} are the mean of the runs $N=1\rightarrow 15$, measured on an Intel i7-9750H. Model names are Double Vault, Archway, Vault, Torus, and Pavilion.
    }
    \label{tab:results}
    \begin{tabular}{r r r r r r }
      & \multicolumn{1}{c}{D.V.} & \multicolumn{1}{c}{A.W.} & \multicolumn{1}{c}{V.} &  \multicolumn{1}{c}{T.}& \multicolumn{1}{c}{P.}    \\
      \midrule
      $N$      & 5 &  9 &  6 &  12 &   5      \\
      m        & 35 &  135 &  25 &  81 &   81      \\
      $t_{opt}$& 7.5s & 44.4s  & 6.4s & 87.5s &   25.6s   \\[0.1cm]
      $E_{\text{ass}}$ & \multicolumn{1}{r}{{0.0086}} & \multicolumn{1}{r}{{0.0098}} & \multicolumn{1}{r}{{0.0092}} & \multicolumn{1}{r}{{0.0050}} & \multicolumn{1}{r}{{0.0086}} \\
    \end{tabular}
  \end{minipage}\hfill%
  \begin{minipage}[t]{0.49\textwidth}
  \vspace{0pt}
    \centering     \includegraphics[width=\columnwidth]{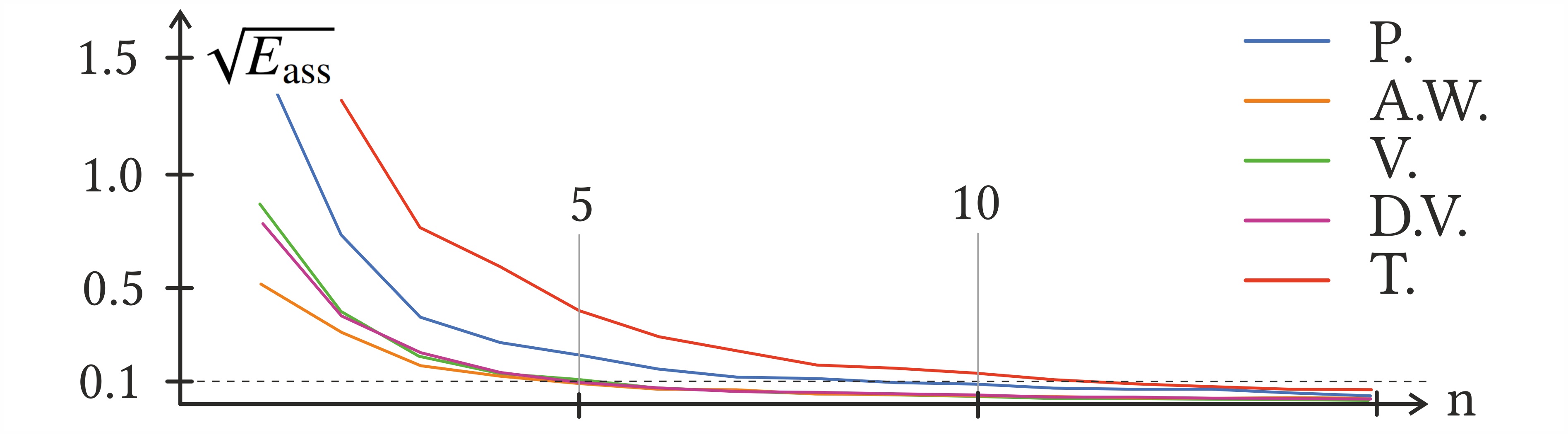}
    \caption{Decrease in our geometric energy $E_{ass}$ with increasing number of linear segments $N$. Our target deviation is depicted as a grey dotted line. Please note that $N=2$ corresponds to a polyline with three vertices, as $c(t=0)$ and $c(t=1)$ are always part of $\overline {c}(\mathbf{t})$.}
    \label{fig:diagram}
  \end{minipage}
\end{figure*}

\paragraph{Algorithm}
 {Given the dense inverse-traced trajectories
$\mathcal{C} = \{\mathbf{c}_1(t), \dots, \mathbf{c}_K(t)\}$, we
evaluate prescribed discretization sizes $N$ separately. For each $N$, we
optimize the shared parameter sequence $\mathcal{T}$ by minimizing
$E_{\text{ass}}(\mathcal{T})$ (Eq.~\eqref{eq:opt_assemblies}) via GA,
evaluate $\sqrt{E_{\text{ass}}}$, and select the smallest tested $N$
satisfying $\sqrt{E_{\text{ass}}} < r$, where $r$ is the lamella thickness
(Figure~\ref{fig:diagram}).} The output is a synchronized piecewise-linear
description of the deployment motion {, with shared time steps directly
usable as multi-step displacement boundary conditions in standard FEA
solvers}.  {For a prescribed $N$, the driver performs a single
GA solve over the internal shared parameters $t_1,\ldots,t_{N-1}$
under the monotonicity constraint in Eq.~\eqref{eq:opt_assemblies}; different
discretization sizes are handled as separate runs rather than by an automatic
outer loop inside the GA driver.}

\section{Results} \label{sec:results}

\begin{figure*}[h]
    \centering
    \includegraphics[width=\textwidth]{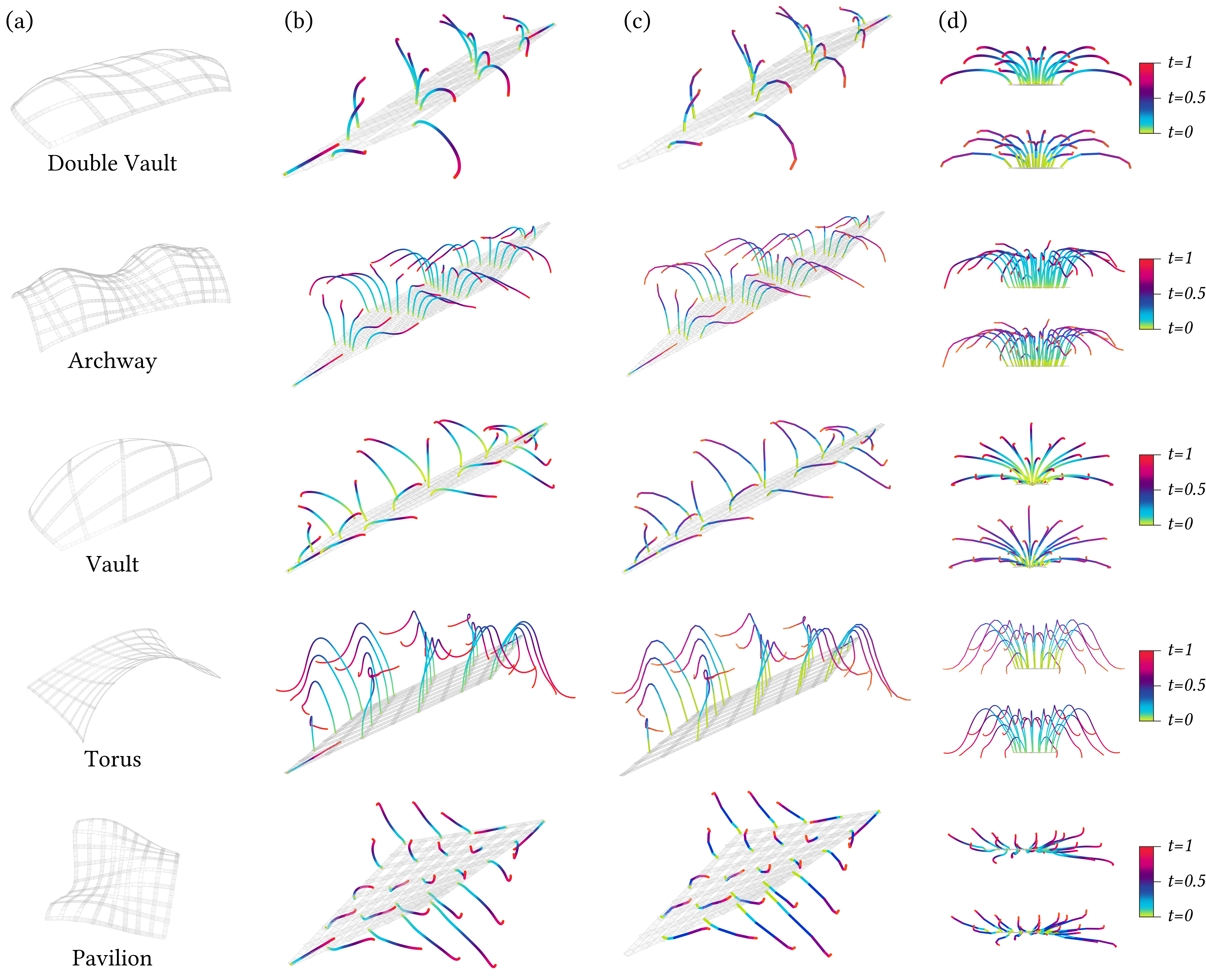}
    \caption{
    Discretization results. 
    (a) Target gridshell models. 
    (b) Subsets of deployment trajectories $c_k(t)$, where color encodes the shared parameter $t$. 
    (c) Synchronized polyline discretizations $\overline{c}_k(\mathbf{t})$. 
    (d) Front view of the deployed grids.
    Numeric metrics are listed in Table \ref{tab:results}.
    } 
    \label{fig:results}
\end{figure*}

We applied our pipeline to the examples depicted in Figure \ref{fig:results} and Table \ref{tab:results}.
The number of linear segments matters for our purposes: a high number allows a better approximation of the curved deployment paths, but we avoid excessive refinement to limit the number of displacement steps.
We set a geometric convergence threshold $\sqrt{E_\text{ass}} < r$, where $r$ is the lamella thickness (Figure \ref{fig:diagram} shows the convergence of $E_\text{ass}$ vs. $N$).  {The metric~$E_{\text{ass}}$ is a tail-aggregated synchronized path-deviation measure over the guided anchor trajectories and thus serves as a direct geometric surrogate for compression risk: a lower~$E_{\text{ass}}$ indicates that the prescribed boundary conditions remain closer to the natural deployment arc, reducing the chord-arc shortfall that generates undesired  axial compression in the slender elements.}

\begin{figure*}
    \centering
    \includegraphics[width=\textwidth]{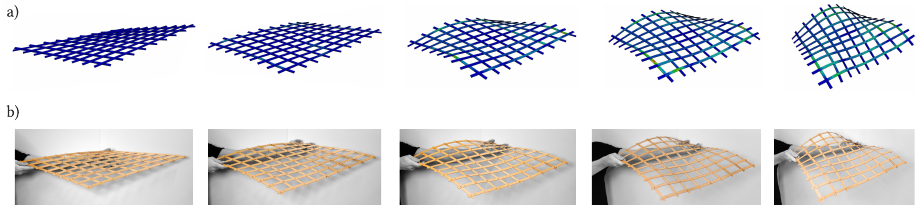}
    \caption{
    Deployment geometry under validation.
    (a) Using synchronized discretization, the finite element deployment simulation proceeds through a sequence of stable steps for a complex gridshell topology \cite{Pillwein2020}.
    (b) The simulated deployed configuration shows close geometric correspondence with the physically fabricated plywood prototype.
    }
    \label{fig:pavilion}
\end{figure*}

\begin{figure*}[t!]
    \centering
    \includegraphics[trim={0.35cm 0cm 0cm 3.8cm},clip, width=\textwidth]{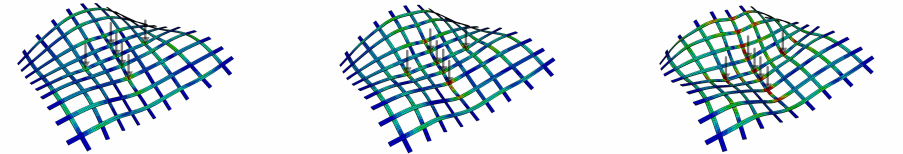}
    \\[10pt]
    \includegraphics[width=\textwidth]{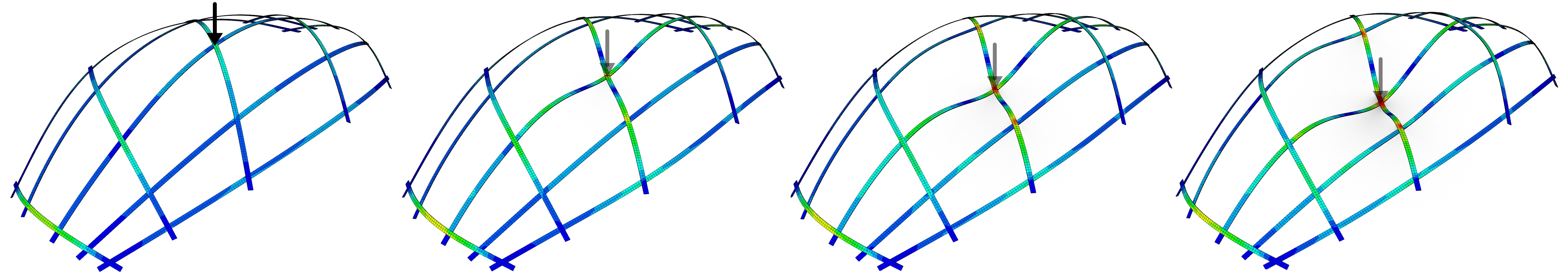}
    \caption{
    Structural response under load.
    The top row shows the qualitative von Mises stress distribution in the deployed configuration, including deployment-induced prestress.
    The subsequent rows show the response of the prestressed grid to a vertical load of increasing magnitude.
    }
\label{fig:snail_loads}
\end{figure*}

\paragraph{Validation via FEA}
We tested our approach with two models, shown in Figures \ref{fig:snail_loads} and \ref{fig:pavilion}: a simple dome shape and a "Pavilion" with multiple curvature regions. Both employ the sliding connections described in ~\ref{sec:abaqus}. We used six boundary points for the dome and sixteen for the complex shape to define the synchronized multi-step paths. 
The FEA simulations (in Abaqus 2019) converged successfully, confirming that the discretized paths avoid buckling-inducing compression. For reference, naive single-step linear actuation ($N=1$) produces the type of buckled equilibria illustrated in Figure~\ref{fig:deployment_fail}; in our Abaqus validation models, the multi-step synchronized approach was required in every case to obtain a valid deployed configuration.

Crucially, the simulation captures the deployment-induced prestress. We compared the structural response of the deployed Pavilion model (approx. 2\% deployment strain) against a stress-free initialization of the same geometry. Under a uniform vertical load, the prestressed model exhibited approximately 8\% less vertical deflection. This agrees with the mechanics of active-bending structures and demonstrates that capturing this prestress via accurate geometric guidance is relevant for structural analysis.

\section{Discussion and Limitations}
\label{sec:discussion}

\paragraph{Multi-step Discretization}
Across our tests, single-step linear actuation ($N=1$) consistently produced geometric deviations and axial compression exceeding buckling tolerances. Introducing synchronized multi-step discretization reduced these deviations, enabling stable deployment. Within the scope of our experiments, multi-step discretization is therefore required to control geometric deviation.

\paragraph{Adaptive Discretization}
Currently, the number of linear segments $N$ is chosen globally to satisfy a deviation threshold. This does not account for local curvature variations. Adaptive refinement strategies based on local nonlinearity could reduce the required steps while maintaining deviation control.

\paragraph{Path-Reversibility Assumption}
 {Inverse tracing is valid when the quasi-static relaxation from
the deployed to the flat state is monotonic in the elastic energy and
stays on a single equilibrium branch. In that regime, the traced path
and the forward deployment trajectory coincide up to reversal of the
time parameter. The assumption fails for structures with pronounced
bistability or snap-through, where the deployed and flat states are
separated by an energy barrier or lie on distinct branches connected
by hysteresis; in that case inverse and forward paths no longer agree
and the derived boundary conditions lose their physical interpretation.
In all models considered here---Double Vault, Archway, Vault, Torus,
and Pavilion---the relaxation stays on a single stable branch
(Section~\ref{sec:results}), so the derived boundary conditions retain
their physical interpretation. Extending the method to bistable or
snap-through systems would require a forward-deployment solver with
explicit branch tracking (e.g.\ arc-length continuation), which is
outside the scope of the present geometric formulation.}

\paragraph{Connection Idealization}
We model sliding connections as idealized kinematic constraints (\ref{sec:abaqus}). This abstraction captures global kinematics but neglects friction or local hole geometry. These effects can be incorporated in detailed finite element models once geometric paths are established.

\paragraph{Material Behavior and Assembly Conditions.}
Our focus is on geometric kinematics. Nonlinear material effects such as creep are not explicitly modeled during path generation and are treated as secondary. Additionally, while the computed trajectories provide a numerically consistent path for validation, real-world assembly may deviate due to scaffolding or manual handling. Thus, geometric guidance validates the mechanism design rather than prescribing on-site actuation.

\section{Conclusions}
\label{sec:conclusions}

We presented a geometry-driven approach for representing and discretizing the deployment of scissor-like elastic grids. The central contribution is a geometric guidance pipeline that constructs globally synchronized, solver-compatible displacement sequences from continuous deployment trajectories. By treating deployment as a coupled path discretization problem, the method provides a geometric representation that bridges design-stage kinematics and numerical simulation.

The approach acquires admissible deployment paths via inverse tracing using a reduced-order rod model and discretizes them through a  {globally synchronized} polyline approximation {---a structural requirement imposed by the one-degree-of-freedom kinematic coupling of the grid---}that controls geometric deviation under a shared parameterization. The resulting displacement sequences can be prescribed directly as boundary conditions, enabling quasi-static deployment simulations without introducing excessive compression or unintended buckling under the discretization settings used.

We evaluated the method on several elastic geodesic grid examples and validated the geometric representation by applying it in finite element deployment simulations. These simulations demonstrate that explicitly accounting for the deployment path allows structural analyses to incorporate deployment-induced prestress and associated nonlinear effects. Overall, the proposed geometric guidance provides a principled representation of deployment trajectories that supports reliable simulation and validation of deployable elastic grid structures.

{
\small
\bibliographystyle{./model2-names}\biboptions{authoryear}
\bibliography{related}

@book{Goldberg1989,

author = {Goldberg, David E.},
title = {Genetic Algorithms in Search, Optimization and Machine Learning},
year = {1989},
isbn = {0201157675},
publisher = {Addison-Wesley Longman Publishing Co., Inc.},
address = {USA},
edition = {1st},
}

@article{Chen21,

author = {Chen, Tian and Panetta, Julian and Schnaubelt, Max and Pauly, Mark},
title = {Bistable Auxetic Surface Structures},
year = {2021},
issue_date = {August 2021},
publisher = {Association for Computing Machinery},
address = {New York, NY, USA},
volume = {40},
number = {4},
pages = {39:1--39:9},
issn = {0730-0301},
doi = {10.1145/3450626.3459940},
journal = {ACM Trans. Graph.},
month = {jul},
articleno = {39},
numpages = {9},
}

@article{Rabinovich19,

author = {Rabinovich, Michael and Hoffmann, Tim and Sorkine-Hornung, Olga},
title = {Modeling Curved Folding with Freeform Deformations},
year = {2019},
issue_date = {December 2019},
publisher = {Association for Computing Machinery},
address = {New York, NY, USA},
volume = {38},
number = {6},
issn = {0730-0301},
doi = {10.1145/3355089.3356531},
journal = {ACM Trans. Graph.},
month = {nov},
articleno = {170},
numpages = {12},
}

@article{Binninger21,
journal = {Computer Graphics Forum (proceedings of SGP 2021)},
title = {Developable Approximation via Gauss Image Thinning},
author = {Binninger, Alexandre and Verhoeven, Floor and Herholz, Philipp and Sorkine-Hornung, Olga},
year = {2021},
volume = {40},
number = {5},
pages = {289-300},
issn = {1467-8659},
doi = {10.1111/cgf.14374},
}

@article{Stein18,

author = {Stein, Oded and Grinspun, Eitan and Crane, Keenan},
title = {Developability of Triangle Meshes},
year = {2018},
issue_date = {August 2018},
publisher = {Association for Computing Machinery},
address = {New York, NY, USA},
volume = {37},
number = {4},
pages = {77:1--77:14},
issn = {0730-0301},
doi = {10.1145/3197517.3201303},
journal = {ACM Trans. Graph.},
month = {jul},
articleno = {77},
numpages = {14},
}

@article{Jiang20,

author = {Jiang, Caigui and Wang, Cheng and Rist, Florian and Wallner, Johannes and Pottmann, Helmut},
title = {Quad-Mesh Based Isometric Mappings and Developable Surfaces},
year = {2020},
issue_date = {August 2020},
publisher = {Association for Computing Machinery},
address = {New York, NY, USA},
volume = {39},
number = {4},
pages = {128:1--128:13},
issn = {0730-0301},
doi = {10.1145/3386569.3392430},
journal = {ACM Trans. Graph.},
month = {jul},
articleno = {128},
numpages = {13},
}

@article{Tang16,

author = {Tang, Chengcheng and Bo, Pengbo and Wallner, Johannes and Pottmann, Helmut},
title = {Interactive Design of Developable Surfaces},
year = {2016},
issue_date = {May 2016},
publisher = {Association for Computing Machinery},
address = {New York, NY, USA},
volume = {35},
number = {2},
pages = {12:1--12:12},
issn = {0730-0301},
doi = {10.1145/2832906},
journal = {ACM Trans. Graph.},
month = {jan},
articleno = {12},
numpages = {12},
}

@article{Ren21,

author = {Ren, Yingying and Panetta, Julian and Chen, Tian and Isvoranu, Florin and Poincloux, Samuel and Brandt, Christopher and Martin, Alison and Pauly, Mark},
title = {3D Weaving with Curved Ribbons},
year = {2021},
issue_date = {August 2021},
publisher = {Association for Computing Machinery},
address = {New York, NY, USA},
volume = {40},
number = {4},
pages = {127:1--127:15},
issn = {0730-0301},
doi = {10.1145/3450626.3459788},
journal = {ACM Trans. Graph.},
month = {jul},
articleno = {127},
numpages = {15},
}

@article{Haskell21,
author = {Charles Haskell and Nicolas Montagne and Cyril Douthe and Olivier Baverel and Corentin Fivet},
title ={Generation of elastic geodesic gridshells with anisotropic cross sections},
journal = {International Journal of Space Structures},
volume = {36},
number = {4},
pages = {294-306},
year = {2021},
doi = {10.1177/09560599211064099}}

@article{Pillwein2021b,

author = {Pillwein, Stefan and Musialski, Przemyslaw},
title = {{Generalized Deployable Elastic Geodesic Grids}},
year = {2021},
issue_date = {December 2021},
publisher = {Association for Computing Machinery},
address = {New York, NY, USA},
volume = {40},
number = {6},
pages = {271:1--271:15},
issn = {0730-0301},
doi = {10.1145/3478513.3480516},
journal = {ACM Trans. Graph.},
month = {dec},
articleno = {271},
numpages = {15}, 
archivePrefix = {arXiv},
arxivId = {2111.08883}, 
}

@article{Autengruber2020,

title = {Finite-element-based moisture transport model for wood including free water above the fiber saturation point},
journal = {International Journal of Heat and Mass Transfer},
volume = {161},
pages = {120228},
year = {2020},
issn = {0017-9310},
doi = {10.1016/j.ijheatmasstransfer.2020.120228},
author = {Maximilian Autengruber and Markus Lukacevic and Josef F{\"u}ssl},
}

@article{Autengruber2021,

title = {Finite-element-based concept to predict stiffness, strength, and failure of wood composite I-joist beams under various loads and climatic conditions},
journal = {Engineering Structures},
volume = {245},
pages = {112908},
year = {2021},
issn = {0141-0296},
doi = {10.1016/j.engstruct.2021.112908},
author = {Maximilian Autengruber and Markus Lukacevic and Gregor Wenighofer and Raimund Mauritz and Josef F{\"u}ssl},
}

@article{Sakai2020,

  title={A 3-dimensional elastic beam model for form-finding of bending-active gridshells},
  author={Sakai, Yusuke and Ohsaki, Makoto and Adriaenssens, Sigrid},
  journal={International Journal of Solids and Structures},
  volume={193--194},
  pages={328--337},
  year={2020},
  publisher={Elsevier},
  doi={10.1016/j.ijsolstr.2020.02.034},
}

@article{Panetta2021,

author = {Panetta, Julian and Isvoranu, Florin and Chen, Tian and Si\'{e}fert, Emmanuel and Roman, Beno\^{\i}t and Pauly, Mark},
title = {Computational Inverse Design of Surface-Based Inflatables},
year = {2021},
issue_date = {August 2021},
publisher = {Association for Computing Machinery},
address = {New York, NY, USA},
volume = {40},
number = {4},
pages = {40:1--40:14},
issn = {0730-0301},
doi = {10.1145/3450626.3459789},
journal = {ACM Trans. Graph.},
month = jul,
articleno = {40},
numpages = {14},
}

@article{pillwein2021,

  title={Design and fabrication of multi-patch elastic geodesic grid structures},
  author={Pillwein, Stefan and K{\"u}bert, Johanna and Rist, Florian and Musialski, Przemyslaw},
  journal={Computers \& Graphics},
  volume={98},
  pages={218--230},
  year={2021},
  publisher={Elsevier},
  doi={10.1016/j.cag.2021.06.002},
}

@article{Lefevre2017,

  title={A 4-degree-of-freedom Kirchhoff beam model for the modeling of bending--torsion couplings in active-bending structures},
  author={Lefevre, Baptiste and Tayeb, Fr{\'e}d{\'e}ric and Du Peloux, Lionel and Caron, Jean-Fran{\c{c}}ois},
  journal={International Journal of Space Structures},
  volume={32},
  number={2},
  pages={69--83},
  year={2017},
  publisher={SAGE Publications Sage UK: London, England},
  doi={10.1177/0266351117714346},
}

@article{dAmico2015,

  title={Timber gridshells: Numerical simulation, design and construction of a full scale structure},
  author={D'Amico, B and Kermani, A and Zhang, H and Pugnale, A and Colabella, S and Pone, S},
  journal={Structures},
  volume={3},
  pages={227--235},
  year={2015},
  publisher={Elsevier},
  doi={10.1016/j.istruc.2015.05.002},
}

@article{Guseinov2020,

  title={Programming temporal morphing of self-actuated shells},
  author={Guseinov, Ruslan and McMahan, Connor and P{\'e}rez, Jes{\'u}s and Daraio, Chiara and Bickel, Bernd},
  journal={Nature Communications},
  volume={11},
  number={1},
  pages={1--7},
  doi = {10.1038/s41467-019-14015-2},
  year={2020},
  publisher={Nature Publishing Group},
}

@article{peloux2016,

  title={Construction of a large composite gridshell structure: a lightweight structure made with pultruded glass fibre reinforced polymer tubes},
  author={Du Peloux, Lionel and Tayeb, Fr{\'e}d{\'e}ric and Baverel, Olivier and Caron, Jean-Fran{\c{c}}ois},
  journal={Structural Engineering International},
  volume={26},
  number={2},
  pages={160--167},
  year={2016},
  publisher={Taylor \& Francis},
  doi={10.2749/101686616X14555428758885},
}

@article{lara2018,

  title={Long-term bending stress relaxation in timber laths for the structural design of lattice shells},
  author={Lara-Bocanegra, Antonio Jos{\'e} and Majano-Majano, Almudena and Arriaga, Francisco and Guaita, Manuel},
  journal={Construction and Building Materials},
  volume={193},
  pages={565--575},
  year={2018},
  publisher={Elsevier},
  doi={10.1016/j.conbuildmat.2018.10.224},
}

@article{gronquist2020,

  title={Computational analysis of hygromorphic self-shaping wood gridshell structures},
  author={Gr{\"o}nquist, Philippe and Panchadcharam, Prijanthy and Wood, Dylan and Menges, Achim and R{\"u}ggeberg, Markus and Wittel, Falk K},
  journal={Royal Society Open Science},
  volume={7},
  number={7},
  pages={192210},
  year={2020},
  publisher={The Royal Society},
  doi={10.1098/rsos.192210},
}

@book{adriaenssens2014shell,

  title={Shell Structures for Architecture: Form Finding and Optimization},
  author={Adriaenssens, Sigrid and Block, Philippe and Veenendaal, Diederik and Williams, Chris},
  year={2014},
  publisher={Routledge},
}

@article{Lienhard2013,
author = {Lienhard, Julian and Alpermann, Holger and Gengnagel, Christoph and Knippers, Jan},
doi = {10.1260/0266-3511.28.3-4.187},
issn = {0266-3511},
journal = {International Journal of Space Structures},
month = sep,
number = {3-4},
pages = {187--196},
publisher = {SAGE PublicationsSage UK: London, England},
title = {{Active Bending, a Review on Structures where Bending is Used as a Self-Formation Process}},
volume = {28},
year = {2013},
}

@article{downland,

  title={The structural engineering of the Downland Gridshell},
  author={Harris, R and Kelly, O},
  journal={Space Structures 5},
  volume={1},
  pages={161--172},
  year={2002},
  publisher={Thomas Telford},
  doi={10.1680/ss5v1.31739.0018},
}

@article{Harris2003,
author = {Harris, Richard and Romer, John and Kelly, Oliver and Johnson, Stephen},
doi = {10.1080/0961321032000088007},
issn = {0961-3218},
journal = {Building Research {\&} Information},
month = nov,
number = {6},
pages = {427--454},
publisher = { Taylor {\&} Francis Group },
title = {{Design and construction of the Downland Gridshell}},
volume = {31},
year = {2003},
}

@inproceedings{Pillwein2020a,

author = {Pillwein, Stefan and K\"{u}bert, Johanna and Rist, Florian and Musialski, Przemyslaw},
title = {Design and Fabrication of Elastic Geodesic Grid Structures},
year = {2020},
isbn = {9781450381703},
publisher = {Association for Computing Machinery},
address = {New York, NY, USA},
doi = {10.1145/3424630.3425412},
booktitle = {Symposium on Computational Fabrication},
articleno = {2},
numpages = {11},
location = {Virtual Event, USA},
series = {SCF '20},
pages = {2:1--2:11},
}

@inproceedings{laccone2019flexmaps,
  title={FlexMaps Pavilion: a twisted arc made of mesostructured flat flexible panels},
  author={Laccone, Francesco and Malomo, Luigi and P{\'E}rez, Jes{\`u}s and Pietroni, Nico and Ponchio, Federico and Bickel, Bernd and Cignoni, Paolo},
  booktitle={Proceedings of IASS Annual Symposia},
  volume={2019},
  number={5},
  pages={1--7},
  year={2019},
  organization={International Association for Shell and Spatial Structures (IASS)},
}

@inproceedings{X_shells_pavilion,

  title={X-Shell Pavilion: A Deployable Elastic Rod Structure},
  author={Isvoranu, Florin and Panetta, Julian and Chen, Tian and Bouleau, Etienne and Pauly, Mark},
  booktitle={Proceedings of IASS Annual Symposia},
  volume={2019},
  number={5},
  pages={1--8},
  year={2019},
  organization={International Association for Shell and Spatial Structures (IASS)},
}

@article{Pillwein2020,

author = {Pillwein, Stefan and Leimer, Kurt and Birsak, Michael and Musialski, Przemyslaw},
title = {{On Elastic Geodesic Grids and Their Planar to Spatial Deployment}},
doi = {10.1145/3386569.3392490},
journal = {ACM Trans. Graph.},
number = {4},
volume = {39},
year = {2020},
pages = {125:1--125:12},
month = jul,
}

@article{Panetta2019,

author = {Panetta, Julian and Konakovi{\'{c}}-Lukovi{\'{c}}, Mina and Isvoranu, Florin and Bouleau, Etienne and Pauly, Mark},
doi = {10.1145/3306346.3323040},
issn = {0730-0301},
journal = {ACM Trans. Graph.},
month = jul,
number = {4},
pages = {83:1--83:15},
title = {{X-Shells: a new class of deployable beam structures}},
volume = {38},
year = {2019},
}

@inproceedings{Soriano2019,

address = {Barcelona, Spain},
author = {Soriano, Enrique and Sastre, Ramon and Boixader, Dionis},
booktitle = {IASS Symposium 2019, Structural Membranes 2019: Form and Force},
isbn = {978-84-121101-0-4},
pages = {1894--1901},
publisher = {International Centre for Numerical Methods in Engineering (CIMNE)},
title = {{G-shells: Flat collapsible geodesic mechanisms for gridshells}},
url = {https://hdl.handle.net/2117/330614},
year = {2019},
}

@inproceedings{Schling2018,

author = {Schling, Eike and Kilian, Martin and Wang, Hui and Schikore, Jonas and Pottmann, Helmut},
booktitle = {Advances in Architectural Geometry (AAG) 2018},
pages = {140--165},
publisher = {Klein Publishing GmbH},
title = {{Design and construction of curved support structures with repetitive parameters}},
year = {2018},
url = {https://www.geometrie.tuwien.ac.at/ig/publications/curvedsupport/curvedsupport.html},
}

@article{Happold1975,

author = {Happold, Edmund and Liddell, Ian},
journal = {The Structural Engineer},
number = {3},
title = {{Timber Lattice Roof for the Mannheim Bundesgartenschau}},
volume = {53},
year = {1975},
}

@article{Konakovic2016,

author = {Konakovi{\'{c}}, Mina and Crane, Keenan and Deng, Bailin and Bouaziz, Sofien and Piker, Daniel and Pauly, Mark},
doi = {10.1145/2897824.2925944},
issn = {07300301},
journal = {ACM Trans. Graph.},
month = jul,
number = {4},
pages = {89:1--89:11},
publisher = {ACM},
title = {{Beyond developable}},
volume = {35},
year = {2016},
}

@misc{Shukhov1896,

author = {Shukhov, Vladimir},
title = {{Rotunda of the Panrussian Exposition (Nizhny Novgorod, 1896) | Structurae}},
urldate = {2020-01-19},
year = {1896},
url = {https://structurae.net/en/structures/rotunda-of-the-panrussian-exposition},
}

@article{Bergou2008,

address = {New York, New York, USA},
author = {Bergou, Mikl{\'{o}}s and Wardetzky, Max and Robinson, Stephen and Audoly, Basile and Grinspun, Eitan},
doi = {10.1145/1360612.1360662},
isbn = {9781450302104},
issn = {07300301},
journal = {ACM Trans. Graph.},
month = aug,
number = {3},
pages = {63:1--63:12},
publisher = {ACM Press},
title = {{Discrete elastic rods}},
volume = {27},
year = {2008},
}

@article{Kilian2017,

author = {Kilian, Martin and Monszpart, Aron and Mitra, Niloy J. },
doi = {10.1145/3015460},
issn = {07300301},
journal = {ACM Trans. Graph.},
month = may,
number = {3},
pages = {25:1--25:13},
publisher = {ACM},
title = {{String Actuated Curved Folded Surfaces}},
volume = {36},
year = {2017},
}

@article{Guseinov2017,

author = {Guseinov, Ruslan and Miguel, Eder and Bickel, Bernd},
doi = {10.1145/3072959.3073709},
issn = {07300301},
journal = {ACM Trans. Graph.},
month = jul,
number = {4},
pages = {64:1--64:12},
publisher = {ACM},
title = {{CurveUps}},
volume = {36},
year = {2017},
}

@article{Konakovic-Lukovic2018,

author = {Konakovi{\'{c}}-Lukovi{\'{c}}, Mina and Panetta, Julian and Crane, Keenan and Pauly, Mark},
doi = {10.1145/3197517.3201373},
issn = {07300301},
journal = {ACM Trans. Graph.},
month = jul,
number = {4},
pages = {106:1--106:13},
publisher = {ACM},
title = {{Rapid deployment of curved surfaces via programmable auxetics}},
volume = {37},
year = {2018},
}

@article{Bergou2010,

address = {New York, New York, USA},
author = {Bergou, Mikl{\'{o}}s and Audoly, Basile and Vouga, Etienne and Wardetzky, Max and Grinspun, Eitan},
doi = {10.1145/1778765.1778853},
isbn = {9781450301121},
issn = {07300301},
journal = {ACM Trans. Graph.},
month = jul,
number = {4},
pages = {116:1--116:10},
publisher = {ACM Press},
title = {{Discrete viscous threads}},
volume = {29},
year = {2010},
}

@article{Kilian2008,

address = {New York, New York, USA},
author = {Kilian, Martin and Fl{\"{o}}ry, Simon and Chen, Zhonggui and Mitra, Niloy J. and Sheffer, Alla and Pottmann, Helmut},
doi = {10.1145/1360612.1360674},
isbn = {9781450301121},
issn = {07300301},
journal = {ACM Trans. Graph.},
month = aug,
number = {3},
pages = {75:1--75:9},
publisher = {ACM Press},
title = {{Curved folding}},
volume = {27},
year = {2008},
}

@article{Laccone2021,
   author = {Francesco Laccone and Luigi Malomo and Nico Pietroni and Paolo Cignoni and Tim Schork},
   doi = {10.1016/j.istruc.2021.08.004},
   issn = {2352-0124},
   journal = {Structures},
   month = dec,
   pages = {979--994},
   publisher = {Elsevier},
   title = {Integrated computational framework for the design and fabrication of bending-active structures made from flat sheet material},
   volume = {34},
   year = {2021},
}

@article{Malomo2018a,

author = {Malomo, Luigi and P{\'{e}}rez, Jes{\'{u}}s and Iarussi, Emmanuel and Pietroni, Nico and Miguel, Eder and Cignoni, Paolo and Bickel, Bernd},
doi = {10.1145/3272127.3275076},
issn = {07300301},
journal = {ACM Trans. Graph.},
month = dec,
number = {6},
pages = {241:1--241:14},
publisher = {ACM},
title = {{FlexMaps}},
volume = {37},
year = {2018},
}

@article{Vekhter2019,

author = {Vekhter, Josh and Zhuo, Jiacheng and Fandino, Luisa F Gil and Huang, Qixing and Vouga, Etienne},
doi = {10.1145/3306346.3323043},
issn = {07300301},
journal = {ACM Trans. Graph.},
month = jul,
number = {4},
pages = {34:1--34:22},
title = {{Weaving geodesic foliations}},
volume = {38},
year = {2019},
}
}

\appendix

\section{Asymptotic Suboptimality of Uniform Synchronized Subdivision}
\label{app:uniform-suboptimal}

 {%
We analyze a synchronized minimax chord--arc surrogate, distinct from
$E_{\mathrm{ass}}$, to give the geometric reason why non-uniform
spacing is preferred. The paper-level statement---that uniform spacing
cannot outperform a global minimizer on $E_{\mathrm{ass}}$---follows
directly from Section~\ref{sec:synchronized_discretization}, where
$\mathbf{t}^{\mathrm{u}}$ is one feasible point of
Eq.~\eqref{eq:opt_assemblies}.

\paragraph{Setup}
Let $c_1,\dots,c_m:[0,1]\to\mathbb{R}^3$ be $C^3$ regular deployment
paths with shared parameter $t$. For a subdivision
$\mathbf{t}=(0=t_0<\cdots<t_N=1)$ with $h_i:=t_i-t_{i-1}$, let
$\ell_{i,j}$ be the chord joining $c_j(t_{i-1})$ and $c_j(t_i)$. Define
\begin{align}
E(\mathbf{t}) &:=
\max_{\substack{1\le i\le N\\1\le j\le m}}
\max_{t\in[t_{i-1},t_i]}
\|c_j(t)-\ell_{i,j}(t)\|,
\label{eq:appendix-surrogate}\\
\Phi(t) &:= \tfrac{1}{8}\max_{1\le j\le m}\|c_j''(t)\|,
\qquad S:=\int_0^1\!\sqrt{\Phi(\tau)}\,d\tau,
\label{eq:appendix-envelope}
\end{align}
and assume $\Phi$ positive and non-constant on $[0,1]$.

\paragraph{Proposition}
\emph{Let $\mathbf{t}^{\mathrm{u}}=(i/N)_{i=0}^N$ be the uniform
subdivision and let $\mathbf{t}^*$ be defined by
$\int_0^{t_i^*}\!\sqrt{\Phi}\,d\tau = iS/N$. Then}
\begin{equation}
E(\mathbf{t}^{\mathrm{u}}) = \frac{\max_t\Phi(t)}{N^2}+O(N^{-3}),
\qquad
E(\mathbf{t}^*) = \frac{S^2}{N^2}+O(N^{-3}),
\label{eq:appendix-rates}
\end{equation}
\emph{and $E(\mathbf{t}^*)<E(\mathbf{t}^{\mathrm{u}})$ for all
sufficiently large $N$.}

\paragraph{Proof}
Taylor expansion and $\max_{s\in[0,h_i]}s(h_i-s)=h_i^2/4$ give
\begin{equation}
E(\mathbf{t}) = M(\mathbf{t})+O(h_{\max}^3),\qquad
M(\mathbf{t}):=\max_{i}\Bigl(\sup_{t\in I_i}\Phi(t)\Bigr)h_i^2.
\label{eq:appendix-M}
\end{equation}
For $\mathbf{t}^{\mathrm{u}}$ one interval contains a maximizer of
$\Phi$, so $M(\mathbf{t}^{\mathrm{u}})=\max_t\Phi(t)/N^2$. For
$\mathbf{t}^*$, applying the mean-value theorem to
$\int_{t_{i-1}^*}^{t_i^*}\!\sqrt\Phi\,d\tau=S/N$ gives
$h_i^*=S/(N\sqrt{\Phi(\eta_i)})$ for some $\eta_i\in I_i^*$;
continuity of $\Phi$ then yields
$(\sup_{I_i^*}\Phi)(h_i^*)^2=S^2/N^2+O(N^{-3})$ uniformly in $i$,
hence $M(\mathbf{t}^*)=S^2/N^2+O(N^{-3})$.
Strict inequality $S^2<\max_t\Phi(t)$ follows from
$\int_0^1\!\sqrt{\Phi}\,d\tau<\sqrt{\max_t\Phi(t)}$,
which holds because $\sqrt{\Phi(t)}<\sqrt{\max\Phi}$ on a set of
positive measure when $\Phi$ is continuous and non-constant. \qed

\paragraph{Consequence}
The asymptotic suboptimality ratio
\[
\frac{E(\mathbf{t}^{\mathrm{u}})}{E(\mathbf{t}^*)}
\;\longrightarrow\;
\frac{\max_t \Phi(t)}{S^2} \;>\; 1
\qquad (N\to\infty)
\]
quantifies the gap between uniform and optimal synchronized subdivision
at leading order, and is attained for the equidistributing sequence
$\mathbf{t}^*$ defined above.

Whenever $\Phi$ is non-constant, uniform synchronized subdivision
misallocates steps: it undersamples the high-curvature parts of the
motion and oversamples the flatter ones. The non-uniform subdivision
$\mathbf{t}^*$ corrects this by concentrating steps where $\Phi$ is
large, and therefore attains a strictly smaller worst-case chord--arc
error for all sufficiently large $N$. Uniform spacing is therefore only
the unoptimized limit of the same synchronized discretization problem.

}

\section{Implementation Details}
\label{sec:abaqus}

\subsection{Geometric Connection Formulation}
\label{sec:connection_geometry}

The deployment mechanism relies on sliding connections that possess one rotational and two translational degrees of freedom (DoF). Explicitly modeling the contact mechanics of physical pins and slots is computationally prohibitive for global deployment simulation. Instead, we formulate a kinematic abstraction where the connecting element (pin) is constrained to lie on the centerlines of both intersecting lamellae while defining their center of rotation.

We model the slot limits as line segments parameterized by $t \in [0,1]$. For a connection between two elements, the position of the pin $\mathbf{x}$ relative to the slot endpoints is given by:
\begin{align}
    \mathbf{x}_1(t_1) = \mathbf{p}_{B} t_1  +  \mathbf{p}_{A} (1-t_1) \,,  \label{eq:DoF_trans1} \\
    \mathbf{x}_2(t_2) = \mathbf{p}_{D} t_2  +  \mathbf{p}_{C} (1-t_2) \,,
    \label{eq:DoF_trans2}
\end{align}
where $\mathbf{p}_{A}, \mathbf{p}_{B}$ and $\mathbf{p}_{C}, \mathbf{p}_{D}$ are the endpoints of the slots on element 1 and 2, respectively (Figure \ref{fig:notches_abstract}). The kinematic constraint enforces that the pin locations coincide in world space, coupling the elements:
\begin{equation}
    \| \mathbf{x}_{1}(t_1) - \mathbf{x}_{2}(t_2) \|^2 = 0\,.
    \label{eq:DoF_rot}
\end{equation}
Together with surface normal contact to prevent interpenetration, these constraints define the allowable kinematic subspace of the connection: rotation around the common normal and translation within the parametric bounds $[0,1]$.

\subsection{Finite Element Instantiation}
\label{sec:abaqus_instantiation}

\begin{figure*}[t]
  \centering
  \begin{minipage}[t]{0.49\textwidth}
    \centering
    \includegraphics[width=\columnwidth]{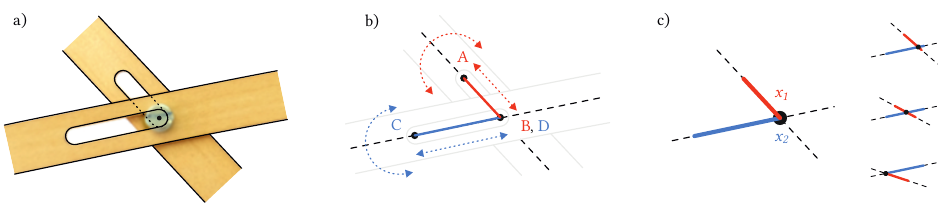}
    \caption{
    Kinematic abstraction of the sliding connection.
    (a) Physical geometry with slot and pin.
    (b) Geometric reduction to line segments and a connection point.
    (c) The coupling constraints allow translation $t_{1,2}$ along the segments and rotation $\phi$ around the common normal.
    }
    \label{fig:notches_abstract}
  \end{minipage}\hfill%
  \begin{minipage}[t]{0.49\textwidth}
    \centering
    \includegraphics[width=\columnwidth]{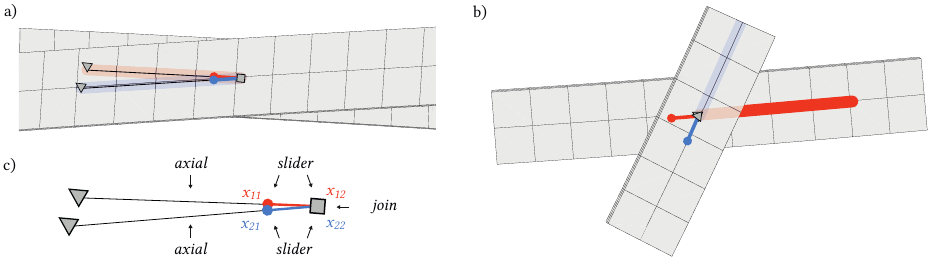}
    \caption{
    Finite element realization.
    (a) Initial configuration: \textit{Wire} elements (colored lines) represent the pins.
    (b) Deployed configuration: Pins have translated to the slot limits.
    (c) Constraint topology: \textit{Slider} MPCs restrict nodes to slot lines; \textit{Join} connectors couple the layers; \textit{Axial} connectors enforce slot length limits.
    }
    \label{fig:notch_abaqus}
  \end{minipage}
\end{figure*}

We implemented this formulation in Abaqus/Standard 2019 using a combination of constraints and connector elements (Figure \ref{fig:notch_abaqus}).

\paragraph{Mesh and Elements}
Lamellae were meshed using C3D10 quadratic tetrahedral elements (2 layers through thickness) to capture bending accurately without hourglassing. The connection pin was modeled as a high-stiffness, two-node \textit{wire} element initialized at the intersection point.

\paragraph{Constraint Network}
The kinematic conditions \eqref{eq:DoF_trans1}--\eqref{eq:DoF_rot} were realized as follows: 
\textbf{Slot Constraint:} The Multi-Point Constraint (MPC) type \textit{Slider} was used to constrain the wire nodes to remain on the lines defined by the slot endpoints. 
\textbf{Pin Coupling:} A \textit{Join} connector coupled the wire nodes of intersecting lamellae, enforcing translational equality ($\|\mathbf{x}_1 - \mathbf{x}_2\| = 0$) while allowing free relative rotation. 
\textbf{Slot Limits:} An \textit{Axial} connector with defined upper/lower bounds enforced the parametric limits $t \in [0,1]$, preventing the pin from sliding off the element.

\paragraph{Solver Settings}
We employed the \textit{General Contact} algorithm with a "hard" normal contact behavior to resolve lamella-to-lamella collisions. The deployment simulation utilized the static solver with automatic time stepping and numerical damping to stabilize rigid body modes during the initial low-stiffness phase.

\begin{table}[h]
	\footnotesize
	\centering
	\caption{
	Simulation statistics. DoF$_\text{user}$ indicates mesh degrees of freedom; DoF$_\text{internal}$ includes Lagrange multipliers for contact/constraints. Run times are for full deployment and loading sequences.
	}
	\label{tab:abaqus}
\begin{tabular}{r r r r r }
	& \multicolumn{1}{c}{DoF$_\text{user}$} &  \multicolumn{1}{c}{DoF$_\text{internal}$} & \multicolumn{1}{c}{$t_\text{deploy}$} & \multicolumn{1}{c}{$t_\text{load}$}    \\
	\midrule
	Pavilion 	& 15334 &  169213 &  9.3h &  3.6h       \\
	Vault 	    & 13681 &  151054 &  7.8h &  2.0h       \\
\end{tabular}
\end{table}

\end{document}